\documentclass[onecolumn,draft,prd,nofootinbib,showkeys]{revtex4}
\usepackage{epsf}
\usepackage{dcolumn}
\usepackage{bm}

\newcommand{\be}{\begin{equation}}
\newcommand{\ee}{\end{equation}}

\def\eg{{\it e.g.},~}
\def\etal{{\it et al.}~}
\def\4he{$^4$He}
\def\3he{$^3$He}
\def\7li{$^7$Li}

\newcommand\la{\lower0.6ex\vbox{\hbox{\ensuremath{\buildrel{\textstyle<}\over{\sim}\ }}}}
\newcommand\ga{\lower0.6ex\vbox{\hbox{\ensuremath{\buildrel{\textstyle>}\over{\sim}\ }}}}

\begin{document}

\title{When Clusters Collide: Constraints On Antimatter On The Largest Scales}

\author{Gary Steigman\\}

\affiliation{Departments of Physics and Astronomy and CCAPP, 
The Ohio State University, Columbus, OH 43210\\}

\date{{\today}\\}

\begin{abstract}

Observations have ruled out the presence of significant amounts of 
antimatter in the Universe on scales ranging from the solar system, to 
the Galaxy, to groups and clusters of galaxies, and even to distances 
comparable to the scale of the present horizon.  Except for the 
model-dependent constraints on the largest scales, the most significant 
upper limits to diffuse antimatter in the Universe are those on the 
$\sim$~Mpc scale of clusters of galaxies provided by the EGRET upper 
bounds to annihilation gamma-rays from galaxy clusters whose intracluster 
gas is revealed through its x-ray emission.  On the scale of individual 
clusters of galaxies the upper bounds to the fraction of mixed matter 
and antimatter for the 55 clusters from a flux-limited x-ray survey 
range from $5 \times 10^{-9}$ to $<  1 \times 10^{-6}$, strongly 
suggesting that individual clusters of galaxies are made entirely of 
matter or, of antimatter.  X-ray and gamma-ray observations of colliding 
clusters of galaxies, such as the Bullet Cluster, permit these constraints 
to be extended to even larger scales.  If the observations of the Bullet 
Cluster, where the upper bound to the antimatter fraction is found to 
be $< 3 \times 10^{-6}$, can be generalized to other colliding clusters 
of galaxies, cosmologically significant amounts of antimatter will be 
excluded on scales of order $\sim 20$ Mpc ($M \sim 5 \times 10^{15}M_{\odot}$).
 
\end{abstract}

\pacs{}

\keywords{Antimatter, X-Ray Clusters, Gamma-Rays, Bullet Cluster}

\maketitle

\section{~Introduction}
\label{intro}

Had the Universe had been matter-antimatter (baryon-antibaryon) 
symmetric during the later stages of its early evolution, when the 
temperature was below that of the quark-hadron transition (and well 
below the nucleon mass), it would have experienced an ``annihilation 
catastrophe" in the sense that the number of post-annihilation 
nucleons would be a billion -- or more -- times less abundant than 
is observed in our present Universe (for a review and an extensive 
list of references, see \cite{steig76}).  Even worse, for a symmetric
universe the annihilation which ceased during its early evolution due 
to the low density of the surviving nucleon-antinucleon pairs, would 
resume when gravitationally-collapsed objects formed (if, indeed, 
collapsed objects could form in such a baryon-poor universe), 
further reducing the baryon density and inhibiting the formation of 
stars, planets, etc.  The problems of a symmetric Universe were 
appreciated by Sakharov~\cite{sakharov}, who outlined the necessary 
conditions for generating a matter-antimatter asymmetry during the 
early evolution of the Universe.  Later, when it was understood 
that Grand Unified Theories, along with the expected, early evolution 
of the standard, hot big bang cosmological model, contained the 
ingredients of Sakharov's recipe for generating a baryon asymmetry 
\cite{yoshimura,susskind,toussaint,weinberg}, it became generally 
accepted that our Universe is matter-antimatter asymmetric, consisting
predominantly of ordinary matter (by definition!) and containing, 
at most, only trace amounts of antimatter.  For a recent, contrary 
point of view, see \cite{bambi1,bambi2,dolgov}.  Over the years, the 
theoretical expectations of a baryon asymmetric Universe have been 
tested and confirmed by a wide variety of observations which strongly 
limit the observationally allowed amount of cosmological antimatter 
\cite{steig76,wolfendale,cdg,bambi1,bambi2,dolgov}.

During the collapse of stars and, in particular, of the pre-solar 
nebula, any relic antimatter would have annihilated to extremely 
low levels.  Even so, it is useful to note that the absence of 
annihilation gamma rays, which could have been produced as the 
solar wind sweeps over the planets, strongly restricts the presence 
of significant amounts of antimatter in the solar system \cite{steig76}.  
Beyond the solar system, only the cosmic rays provide direct evidence 
of the composition of the stars and gas in the Galaxy.  The absence 
of complex antinuclei (\eg antihelium) in the cosmic rays at a level 
$< 10^{-6}$ \cite{ams1} provides an interesting upper bound to antimatter 
in the Galaxy.  New cosmic ray experiments such as PAMELA \cite{pamela} 
and AMS \cite{ams2} have the potential to reduce this limit further or, 
to find evidence for antimatter from observations of galactic or 
extragalactic cosmic ray antinuclei.  While a collapsed object, 
such as a star, may hide appreciable amounts of antimatter\footnote{In 
its journey through the Galaxy, an antistar would accrete interstellar 
gas, leading to the production of annihilation gamma-rays.  Observations 
of discrete Galactic gamma-ray sources limit the fraction of antistars 
in the Galaxy to $< 10^{-4}$ \cite{steig76}.}, stars are formed from 
interstellar gas and, in the course of their evolution and, at the 
end of their evolution, they return substantial amounts of material 
to the interstellar medium.  Since an antinucleon (or, antinucleus) in 
the diffuse interstellar medium (ISM) of the Galaxy has a very short 
lifetime against annihilation, $\sim 300$~yr -- 200 kyr \cite{steig76}, 
any antimatter present when the Galaxy formed wouldn't have survived 
to the present epoch\footnote{The $\sim 3 - 5$~order of magnitude longer 
lifetime in Bambi and Dolgov \cite{bambi1}, while still very small 
compared to the age of the Galaxy, fails to account for the $\sim 
3 - 5$ order of magnitude enhancement of the annihilation cross section 
at the very low collision energies in the ISM.  As a consequence, the 
Bambi and Dolgov gamma-ray flux estimates \cite{bambi1} may need to 
be corrected upwards by 3 -- 5 orders of magnitude.}.  Indeed, the 
observed Galactic gamma rays indirectly limit the ratio of antimatter 
to matter in the ISM to $< 10^{-15}$ \cite{steig76}.  On scales larger 
than that of the Galaxy, upper bounds to the observed gamma-ray flux 
provide indirect limits on the presence of diffuse regions of mixed 
matter and antimatter.

In \S\ref{clusters}, the gamma-ray limits on matter-antimatter annihilation 
in the hot, x-ray emitting gas of clusters of galaxies are reviewed, 
leading to bounds on the antimatter fraction in systems of size 
$\sim$~few Mpc and mass $\sim 10^{15}M_{\odot}$.  In \S\ref{bullet} it 
is noted that observations of the ``Bullet Cluster" \cite{markevitch} 
and of other colliding clusters of galaxies \cite{jee,mahdavi,bradac} 
permit these bounds to be extended to larger distance/mass scales.  
The results presented here are summarized in \S\ref{conclusions}.
   
\section{~Antimatter Constraints On The Scale Of Clusters Of Galaxies}
\label{clusters}

In clusters of galaxies most of the baryons (matter) are in the hot, 
x-ray emitting, intracluster gas.  If a fraction, $f$, of this gas were 
to consist of antibaryons (antimatter) mixed with the dominant baryons 
(or, vice-versa), then the two-body collisions responsible for creating 
the x-rays via thermal bremsstrahlung emission, would ensure the 
production of high-energy gamma rays from matter-antimatter annihilation 
\cite{steig76}.  As a result, the predicted annihilation gamma-ray flux 
is directly tied to -- proportional to -- the observed x-ray flux.  The 
absence of observed gamma-rays bounds the fraction of mixed matter and 
antimatter in the intracluster gas \cite{steig76}.  The best constraints to 
the presence of antimatter on some of the largest scales in the Universe 
($M \sim 10^{14} - 10^{15}~M_{\odot}$; $R \sim$~few Mpc) are provided 
by a comparison of the upper bounds to the cluster $\gamma$-ray flux, 
$F_{\gamma} \equiv F_{\gamma}(> 100$~MeV) photons~cm$^{-2}$~s$^{-1}$, 
to the observed cluster x-ray flux, $F_{\rm X} \equiv F_{\rm X}(2 - 
10$~keV)~erg~cm$^{-2}$~s$^{-1}$ \cite{steig76}.  For a cluster at a distance 
$R$, whose intra-cluster gas fills a volume $V$ and is at a temperature 
$T_{8} \equiv T/10^{8}$K, the x-ray and the annihilation-predicted 
gamma-ray fluxes are \cite{steig76}\footnote{Note that the numerical coefficient 
in eq.~2 differs from that in ref.~\cite{steig76}, due to a more careful 
accounting of the temperature dependent enhancement of the low energy 
annihilation cross section.}
\be
F_{\rm X} = 1.4\times 10^{-23}T_{8}^{1/2}\int{{n_{\rm B}^{2}dV
\over 4\pi R^{2}}}
\ee
and,
\be
F_{\gamma} = 5.4\times 10^{-14}{f \over T_{8}^{1/2}}\int{{n_{\rm B}^{2}dV
\over 4\pi R^{2}}}.
\ee
Since not all the observed $\gamma$-rays will have been produced by 
annihilation, the ratio of $F_{\gamma}$ to $F_{\rm X}$ provides an 
upper bound to $f$,
\be
f \leq 2.6\times 10^{-10}T_{8}({F_{\gamma} \over F_{\rm X}}) = 
3.0\times 10^{-8}T_{\rm keV}({10^{8}F_{\gamma} \over 10^{11}F_{\rm X}}),
\ee
where $T_{\rm keV} \equiv kT$ measured in keV ($T_{\rm keV} = 8.6T_{8}$).

\begin{figure}[htbp]
\centerline
{
\epsfxsize=3.6truein
\epsfbox{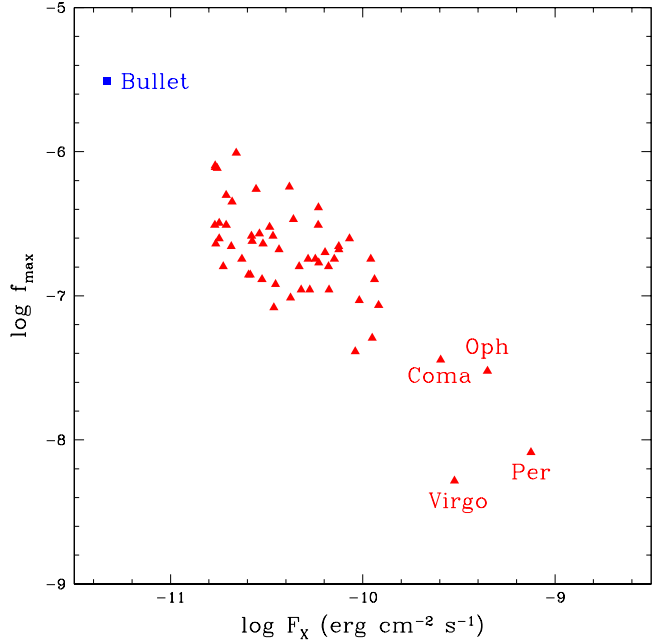}
}
\caption{
The upper limits to the galaxy cluster antimatter fractions, $f_{\rm max}$, 
inferred from the absence of $\gamma$-rays \cite{reimer}, as a function 
of the observed x-ray fluxes \cite{edge}, $F_{\rm X}$, for 55 clusters of 
galaxies (red triangles) and for the Bullet Cluster \cite{markevitch,hartman} 
(blue square).}
\label{fig:fvsfx}
\end{figure}

The flux limited x-ray survey by Edge \etal \cite{edge} identifies 55
clusters emitting at a level $F_{\rm X} \geq 1.7 \times 10^{-11}$ 
erg~cm$^{-2}$~s$^{-1}$.  The upper bounds to the cluster antimatter 
fraction, $f_{\rm max}$, which follow from the EGRET upper bounds to 
the $\gamma$-ray flux \cite{reimer} for these 55 x-ray clusters are shown 
by the red triangles in Figure \ref{fig:fvsfx}, along with the corresponding 
upper bound (blue square) inferred from observations of the Bullet Cluster 
(see \S\ref{bullet}).  These observations limit the fraction of mixed 
matter and antimatter on the scale of clusters of galaxies to be smaller 
than $f < 1 \times 10^{-6}$.  The best constraints (the smallest upper 
bounds) to the antimatter fraction on the scale of galaxy clusters 
are from observations of the Perseus and Virgo clusters, where $M \sim 
10^{15}h_{50}^{-1}M_{\odot}$.  For Perseus, $f < 8\times 10^{-9}$ and,
for Virgo an even lower upper bound of $f < 5\times 10^{-9}$ is derived.  
For the slightly larger scales of the Coma and Ophiucus clusters ($M 
\sim 2 - 3 \times 10^{15}h_{50}^{-1}M_{\odot}$), the observations lead 
to the somewhat weaker, but still very strong, constraints $f < 3 - 
4\times 10^{-8}$.  Indeed, since $f \ll 1$ for all 55 clusters, each of 
these clusters likely consists entirely of matter (or, of antimatter).  
If there are significant antimatter-dominated regions in the Universe, 
they must be separated from matter-dominated regions on scales greater 
than the $\sim$~Mpc ($M \sim 10^{15}h_{50}^{-1}M_{\odot}$) scale of 
clusters of galaxies.  It is of interest to extend the cluster bounds 
to these larger scales.  

\section{~The Bullet Cluster} 
\label{bullet}

It is clear from the data provided by the x-ray and $\gamma$-ray observations 
of a large sample of galaxy clusters \cite{edge,reimer} that if regions of 
diffuse antimatter do exist on large scales in the Universe, they must 
be separated from regions of ordinary matter by distances at least of 
order of the $\sim$~Mpc sizes of clusters of galaxies.  Can the above 
constraints from the x-ray and $\gamma$-ray observations of galaxy clusters 
be extended to even larger distance/mass scales?  For example, how would 
it be known if there were significant amounts of diffuse antimatter in 
regions where the fraction of ordinary matter is very small (\eg entire 
galaxy clusters of antimatter)?  Such regions would reveal themselves 
through the annihilation $\gamma$-rays when the matter and antimatter 
clusters collide.  The only way to probe this possibility of separated 
clusters and anti-clusters is to search for correlated x-rays and 
$\gamma$-rays from colliding clusters of galaxies.  Observations of 
the so-called ``Bullet Cluster'' \cite{markevitch} provide just such 
an opportunity.  

According to Nusser (A. Nusser, Private Communication and \cite{nusser}), 
$M_{\rm Bullet} \sim 6\times 10^{15}h_{50}^{-1}M_{\odot}$ and, at 
maximum, the colliding clusters were separated by $\sim 20$~Mpc.  
For the Bullet cluster, $F_{\rm X} = 4.7\times 10^{-12}$~erg~cm$
^{-2}$~s$^{-1}$, $T_{\rm keV} = 14$ \cite{markevitch} and, at 95\% 
confidence, $F_{\gamma} < 3.5\times 10^{-8}$~photons~cm$^{-2}$~s$^{-1}$ 
(D. Thompson, Private Communication and \cite{hartman}), so that $f_{\rm 
Bullet} < 3\times 10^{-6}$.  While somewhat weaker than the constraints 
on $f$ from the individual clusters discussed in \S\ref{clusters} above, 
this upper bound to the antimatter fraction shows that the colliding 
galaxy clusters which constitute the Bullet Cluster consist predominantly, 
if not entirely, of matter (or, of antimatter!).  If the Bullet 
Cluster is typical, then recent evidence for other clusters in 
collision \cite{jee,mahdavi,bradac} raises the possibility of further 
extending these constraints on antimatter in the Universe to scales 
of tens of Mpc.

\section{~Conclusions}
\label{conclusions}

Direct observations in the solar system and of the galactic cosmic rays
set stringent constraints on antimatter in our local vicinity.  Gamma
rays produced when matter and antimatter meet and annihilate provide 
indirect evidence for regions of mixed matter and antimatter.  There 
is no evidence for such mixed regions on scales from galaxies, to 
groups and clusters of galaxies, limiting the antimatter fraction 
to $< 1\times 10^{-6}$, or smaller, on scales up to $\sim$~Mpc and $M 
\sim 10^{15}M_{\odot}$.  Recent observations of clusters in collision 
\cite{markevitch,jee,mahdavi,bradac} permit the x-ray cluster bounds 
on the fraction of antimatter in the Universe to be extended to even 
larger scales.  Observations of the colliding galaxy clusters which 
constitute the Bullet Cluster limit the antimatter fraction to 
$f_{\rm Bullet} < 3\times 10^{-6}$, on a scale of $M \sim 6\times 
10^{15}h_{50}^{-1}M_{\odot}$.  If, indeed, there are regions of 
antimatter in the Universe, they must be separated from regions of 
ordinary matter by distances on the order of tens of Mpc (mass scales 
of order $10^{16}M_{\odot}$).  Evidence for galaxy clusters in collision 
suggest that these scales can be probed in the near future.

\acknowledgments 

I thank Jim Hartle for asking the question which stimulated this 
project.  I am grateful to Bill Forman, Neil Gehrels, Adi Nusser, 
Olaf Reimer, and Dave Thompson for engaging in helpful, instructive, 
and informative correspondence, and to John Beacom for discussions 
and for suggestions to improve an earlier version of this manuscript.   
This research is supported at The Ohio State University by a grant 
from the US Department of Energy.


\end{document}